





\tolerance=10000
\input phyzzx.tex

\def\W {{\cal W }}

\def\H {{\cal H }}
\def\hh {{\cal H }}

\def\np{Nucl. Phys.}
\def\pl{Phys. Lett.}

\def\cmp{Commun. Math. Phys.}
\def\intmp{Intern. J. Mod. Phys.}

\def\half{{\textstyle {1 \over 2}}}

\def\IB{\relax{\rm I\kern-.18em B}}
\def\IC{\relax\hbox{$\inbar\kern-.3em{\rm C}$}}
\def\ID{\relax{\rm I\kern-.18em D}}
\def\IE{\relax{\rm I\kern-.18em E}}
\def\IF{\relax{\rm I\kern-.18em F}}
\def\IG{\relax\hbox{$\inbar\kern-.3em{\rm G}$}}
\def\IH{\relax{\rm I\kern-.18em H}}
\def\II{\relax{\rm I\kern-.18em I}}
\def\IK{\relax{\rm I\kern-.18em K}}
\def\IL{\relax{\rm I\kern-.18em L}}
\def\IM{\relax{\rm I\kern-.18em M}}
\def\IN{\relax{\rm I\kern-.18em N}}
\def\IO{\relax\hbox{$\inbar\kern-.3em{\rm O}$}}
\def\IP{\relax{\rm I\kern-.18em P}}
\def\IQ{\relax\hbox{$\inbar\kern-.3em{\rm Q}$}}
\def\IR{\relax{\rm I\kern-.18em R}}

\def\weta{{ \tilde h}}
\def\wet{{  h}}
\def\dm {\partial_{\mu}}

\def\ix {\int\!\!d^2x\;}

\REF\hu{C.M. Hull, \pl\ {\bf 240B} (1990) 110.}
\REF\van{K. Schoutens, A. Sevrin and
P. van Nieuwenhuizen, \pl\ {\bf 243B} (1990) 245.}
 \REF\pop{E. Bergshoeff, C.N. Pope, L.J. Romans, E.
Sezgin, X. Shen and K.S. Stelle, \pl\ {\bf 243B}
(1990) 350.}
\REF\hue{C.M. Hull,   \np\ {\bf 353
B} (1991) 707.}
\REF\huee{C.M. Hull, \pl\ {\bf 259B} (1991) 68  and  \np\ {\bf B364} (1991)
621; A. Mikovi\' c, \pl\ {\bf 260B} (1991) 75.}
 \REF\vann{K. Schoutens, A.
Sevrin and P. van Nieuwenhuizen, Nucl. Phys. {\bf B349}
(1991) 791   and Phys. Lett. {\bf 251B} (1990) 355; E. Bergshoeff,
C.N. Pope and K.S. Stelle, \pl\ {\bf 249B} (1990) 208.}
\REF\wevoo{C.M. Hull, in {\it Strings and Symmetries 1991}, ed. by  N.
Berkovits et al,  World Scientific, Singapore, 1992;
Lectures on W-Gravity, W-Geometry and W-Strings, Trieste Summer
School Lectures 1992, to be published by World Scientific,
Singapore.}
\REF\mik{A. Mikovi\' c,
\pl\ {\bf 278B} (1991) 51.}
\REF\bow{P. Bouwknegt and K. Schoutens, CERN preprint CERN-TH.6583/92,
to appear in Physics Reports.}
\REF\wstr{C.N. Pope, L.J. Romans and K.S. Stelle, \pl\ {\bf 268B}
(1991) 167 and {\bf 269B}
(1991) 287.}
\REF\wit{E. Witten, in \lq Proceedings of the Texas A \& M
Superstring Workshop,
 1990', ed. by R. Arnowitt et al, World Scientific Publishing, Singapore,
1991.}
\REF\wot{A. Bilal, \pl\ {\bf 249B} (1990) 56;
A. Bilal, V.V. Fock and I.I. Kogan, \np\ {\bf B359} (1991) 635.}
\REF\sot{G. Sotkov and M. Stanishkov, \np\ {\bf B356} (1991) 439;
G. Sotkov, M. Stanishkov and C.J. Zhu, \np\ {\bf B356} (1991) 245.}
\REF\bers{ M. Berschadsky and H. Ooguri,
\cmp\ {\bf 126} (1989) 49.}
\REF\difran{P. Di Francesco,  C.  Itzykson and J.B. Zuber,
               Commun. Math. Phys. {\bf 140} (1991) 543.}
\REF\ram{J.M. Figueroa-O'Farrill, S. Stanciu and E. Ramos, Leuven preprint
KUL-TF-92-34.}
\REF\gerr{J.-L. Gervais and Y. Matsuo, \pl\ {\bf B274} 309 (1992) and
Ecole Normale
preprint LPTENS-91-351 (1991); Y. Matsuo, \pl\ {\bf B277} 95 (1992)}
\REF\geo{C.M. Hull,  \pl\ {\bf 269B} (1991) 257.}
\REF\hegeoma{C.M. Hull, \W-Geometry, QMW preprint, QMW-92-6 (1992),
 hep-th/9211113, to be published in
 \cmp.}
 \REF\hegeomb{C.M. Hull, Geometry and  \W-Gravity, QMW preprint, QMW-92-21
(1992),
 hep-th/9301074, to be published in Proceedings
 of the
16th John Hopkins Workshop on Current Problems in Particle Theory, Gothenborg,
1992.}
 \REF\mon{T. Aubin, {\it Non-Linear Analysis on Manifolds. Monge-Amp\` ere
Equations}, Springer Verlag, New York, 1982.}
\REF\pleb{J.F. Plebanski, J. Math. Phys. {\bf 16} (1975) 2395.}
 \REF\winf{I. Bakas, \pl\ {\bf 228B} (1989) 57 and Maryland
preprint MdDP-PP-90-033.}
 \REF\park{Q-Han Park, \pl\ {\bf 236B} (1990) 429, {\bf 238B}
(1990) 287 and {\bf 257B} (1991) 105.}
\REF\zam{A.B. Zamolodchikov,
Teor. Mat. Fiz. {\bf 65} (1985) 1205; V.A. Fateev and S.
Lykanov, \intmp\ {\bf A3} (1988) 507; A. Bilal and J.-L. Gervais, \pl\ {\bf
206B} (1988) 412; \np\ {\bf B314} (1989) 646; \np\ {\bf B318} (1989) 579.}
\REF\hit{U. Lindstrom and M. Ro\v cek, \np\ {\bf B222} (1983) 285;
N. Hitchin, A. Karlhede, U. Lindstrom and M. Ro\v cek, Comm. Math. Phys.
{\bf 108} (1987) 535. }
\REF\spec{S.J. Gates, \np\ {\bf B238} (1984) 349; M. Gunaydin, G. Sierra and
P.K. Townsend, \np\ {\bf 242} (1984) 244; B. de Wit, P.G. Lauwers and A. van
Proeyen, \np\ {\bf B255} (1985) 569;
A.
Strominger, \cmp\ {\bf 133} (1990) 163; L. Castellani, R. D'Auria and S.
Ferrara, \pl\ {\bf 241} (1990) 57 and Class. and Quantum Grav. {\bf 7} (1990)
1767.}
\REF\huprep{C.M. Hull, in preparation.}

 \nopubblock
{\begingroup \tabskip=\hsize minus \hsize
   \baselineskip=1.5\ht\strutbox \topspace-2\baselineskip
   \halign to\hsize{\strut #\hfil\tabskip=0pt\crcr
   {QMW-93-6}\cr   {hep-th/9303071} \cr
   {March 1993}\cr }\endgroup}

\titlepage
\title {\bf THE GEOMETRIC STRUCTURE OF $\W_N$-GRAVITY}
\author {C. M. Hull}
\
\address {Physics Department,
Queen Mary and Westfield College,
\break
Mile End Road, London E1 4NS, United Kingdom.}

\abstract
{The full non-linear structure of the action and transformation
rules  for $\W_N$-gravity coupled to  matter are
obtained from a non-linear truncation of
those for $w_ \infty$ gravity.
The geometry of the construction is discussed, and
it is shown that the defining equations become linear
after
a twistor-like transform. }

 \endpage
\pagenumber=1

\chapter {Introduction}

 Classical \W-gravity theories [\hu-\mik] are higher-spin gauge theories
in two dimensions that result from gauging \W-algebras [\bow], which are
higher-spin extensions of the Virasoro algebra.
One motivation for studying two-dimensional matter coupled to
\W-gravity is that  such systems can be interpreted as   generalisations
of string theory in which the two-dimensional space-time  is regarded as a
world-sheet, in much the same way that matter coupled to ordinary gravity in
two dimensions leads to   conventional  string theory. In particular,
the \W-algebras play a central role in such \W-string theories,
just as the Virasoro algebra plays a central role in string theory.
The actions for \W-gravity coupled to matter have a complicated non-polynomial
dependence on the gauge fields. In the case of gravity,
this non-linear structure is best understood in terms of Riemannian geometry
and  this suggests that
 some higher spin geometry might lead to a better understanding of
\W-gravity.
A number of approaches to the geometry of
\W-gravity
 theories have been
considered [\vann,\wit -\hegeomb].
In [\geo,\hegeoma], the complete non-linear structure of the coupling of a
scalar field
on a world-sheet $M$ to $w_\infty$ gravity was given in terms of a function
$ \tilde  F$
on the cotangent bundle of $M$ that satisfied a certain non-linear
differential equation, which is sometimes referred to as a  Monge-Amp\` ere
equation [\mon]
or as one of Plebanski's equations [\pleb].
Such equations  also  arise   in
the study of $4-D$  self-dual gravity  [\pleb]; other
  connections   between \W-algebras and gravitational instantons, which may
be related,  were described in [\winf,\park].
In particular, it was shown in [\geo,\hegeoma] that the function
$ \tilde  F$ could be interpreted as giving a family of K\" ahler potentials
for Ricci-flat metrics on $\IR^4$, with self-dual curvature.
The purpose of this paper is to extend the results of [\geo,\hegeoma] to the
case of
$\W_N$-gravity; some of the results to be derived here were announced in
[\hegeomb].

It will be shown here that the coupling of a scalar field
to $\W_N$ gravity can be given as a non-linear truncation of the
action for the coupling to $w_\infty$ gravity. The lagrangian is
a function $ \tilde  F$
which, in addition
to satisfying the Monge-Amp\` ere equation, satisfies  an $(N+1)$'th
order non-linear partial differential equation, and it is a non-trivial
fact that this constraint is consistent with the
Monge-Amp\` ere equation.
 This differential constraint can
be interpreted geometrically as a condition on
 the family of self-dual
metrics on $\IR ^4$. For $\W_3$, the $4$'th order differential
equation satisfied by the K\" ahler potential can be written as
$$\mathop{ {R}}\nolimits_{\mu\bar{\nu}\rho\bar{\sigma}}={1\over
2}\mathop{ {G}}\nolimits^{\alpha\bar{\beta}}\left\lbrack
T_{\alpha
\mu\bar{\nu}}T_{\bar{\beta}\bar{\sigma}\rho}+T_{\alpha\mu\bar{\sigma
}}T_{\bar{\beta}\bar{\nu}\rho}+T_{\bar{\beta}\bar{\nu}\mu}T_{\alpha
\rho\bar{\sigma}}+T_{\bar{\beta}\bar{\sigma}\mu}T_{\alpha \rho \bar{\nu
}}\right\rbrack
\eqn\curono$$
where $G_{\mu \bar \nu}$ is the K\" ahler metric and $T_{\mu \nu \bar \rho}$
is a certain third rank tensor that is given in terms
of the K\" ahler potential $K$ by
$T_{\mu \nu \bar \rho}= \partial_ \mu \partial _ \nu \partial_{ \bar \rho}K$
in certain special coordinate systems.
   It is interesting to note
that similar, but distinct, geometrical constraints arise in the study of
\lq special geometry', \ie\ the
geometry of the moduli space of Calabi-Yau manifolds,
 and in the geometry of $N=2$ supersymmetric  gauge multiplets
in $4$ and $5$ dimensions [\spec].
 For $\W_N$ with
 $N>3$, the differential constraint can be written as a restriction
 on the $(N-3)$'th covariant derivative of the curvature tensor.

\subsection{Linearised \W-Gravity}

Before proceeding to the non-linear theories, it will be useful to review
linearised $\W_N$ and $w_\infty$ gravity.
Consider the action
for a free scalar field in two
dimensions
 \foot{Flat two-dimensional space $M_0$
has metric $ds^{2}=\mathop{\eta_{\mu\nu}dx^{\mu}dx^{\nu}=}\nolimits2dzd\bar
{z}$, where
 $z={1\over\sqrt{2}}\left(x^{1}+ix^{2}\right)$, $\bar{z}={1\over\sqrt
{2}}\left(x^{1}-ix^{2}\right)$ are complex coordinates if $M_0$ is Euclidean,
while, if $M_0$ is Lorentzian,
$z={1\over\sqrt{2}}\left(x^{1}+x^{2}\right)$, $ \bar{z}={1\over\sqrt
{2}}\left(x^{1}-x^{2}\right)$ are null real coordinates. $ \partial=
\partial_{z}$ and
$\bar{\partial}=\partial _{\bar{z}}$.}
$$
S_0={1 \over 2} \ix   \dm \phi \partial ^\mu \phi
\eqn\free$$
 This has an infinite number of conserved currents, which include [\pop]
$$W_{n}={1\over n}(\partial\phi)^{n} , \qquad n=2,3,....,N
\eqn\wis$$
 and these satisfy the conservation law $\mathop{\bar \partial
W}\nolimits_{n}=0$.
The current $W_2=T$ is a component of the energy-momentum tensor and generates
a Virasoro algebra.
The  currents \wis\ generate
a current algebra which is a certain classical limit of the $\W_N$ algebra of
[\zam]  for finite $N$, and in the limit $N \rightarrow \infty$, the classical
current algebra becomes the $w_\infty$ algebra [\winf].
Similarly, the   currents
 $\mathop{\overline{W}}\nolimits_{n}={1\over n}(\bar \partial\phi)^{n}$
generate a second copy of the $\W_N$ or $w_\infty$ algebra.

Adding the Noether coupling of the currents $W_n,
\mathop{\overline{W}}\nolimits_{n}$ to
corresponding gauge fields $h_n, \bar h_n$ gives the linearised action
$$S=\int
d^{2}x\left\lbrack  \partial\phi \bar \partial \phi+ \sum^{N}_{n=2}{1\over
n}\left[h_{n}(\partial\phi)^{n}+\mathop{\bar
{h}}\nolimits_{n}(\bar{\partial}\phi)^{n}\right]+O(h^{2})\right\rbrack
\eqn\noeth$$
which is invariant, to lowest order in the gauge fields, under the
transformations
 $$\eqalign{
\delta \phi&=\sum^{N}_{n=2}\left[
\lambda_{n}(z,\bar{z})(\partial \phi)^{n-1}
+\mathop{\bar{\lambda}}\nolimits_{n}(z,\bar{z})(\bar \partial \phi)^{n-1}
\right] \cr
\delta h_{n}&=-2\bar{\partial}\lambda_{n}+O(h),\qquad \
\delta\mathop{\bar
{h}}\nolimits_{n}=-2\partial\mathop{\bar{\lambda}}\nolimits_{n}+O(h)
\cr}
\eqn\tre$$
This gives the linearised action and transformations of $\W_N$ or
(in the $N\rightarrow \infty$ limit) $w_\infty$ gravity. The full
gauge-invariant action and gauge transformations  are non-polynomial in the
gauge fields.

\chapter{Non-Linear $w_\infty$-Gravity}

The non-linear structure of the coupling of a scalar field to $w_\infty$
gravity [\geo,\hegeoma] will now be reviewed. The two-dimensional manifold $M$,
which will
sometimes be referred to as the world-sheet, can   have any topology
and has local coordinates $x^\mu$.
 The action is a non-polynomial
function of $\dm \phi$ and can be written as
 $$S=\int _M d^{2}x\tilde{F}(x,\partial\phi)
\eqn\sis$$
for some $\tilde{F}$, which has the following expansion in $y_\mu =\dm \phi$:
$$\tilde{F}(x,y)=\sum^{\infty}_{n=2}{1\over n} {\tilde{g}}
^{\mu_{1}\mu_{2}...\mu_{n}}_{(n)} (x) y _{\mu_{1}}y_{\mu
_{2}}...y_{\mu_{n}}
\eqn\fis$$
where ${\tilde{g}}
^{\mu_{1}\mu_{2}...\mu_{n}}_{(n)}(x)$ are symmetric tensor (density)
gauge fields.

The  gauge fields $\tilde g_{(2)},\tilde g_{(3)},\tilde g_{(4)},  \dots$
are required to satisfy an infinite set of
constraints, the first few of which are
$$\det\left(\mathop{\tilde{g}}\nolimits^{\mu\nu}_{(2)}\right)=
 \epsilon \eqn\decona$$
$$\mathop{\tilde{g}}\nolimits_{\mu\nu}\mathop{\tilde{g}}\nolimits
^{\mu\nu\rho}_{(3)}=0 \eqn\deconb$$
$$\mathop{\tilde{g}}\nolimits_{\mu\nu}\mathop{\tilde{g}}\nolimits
^{\mu\nu\rho\sigma}_{(4)}={2\over3}\mathop{\tilde{g}}\nolimits_{\mu
\alpha}\mathop{\tilde{g}}\nolimits_{\nu\beta}\mathop{\tilde{g}}\nolimits
^{\mu\beta\rho}_{(3)}\mathop{\tilde{g}}\nolimits^{\nu\alpha\sigma
}_{(3)}\eqn\deconc$$
where
 $\mathop{\tilde{g}}\nolimits_{\mu\nu}$ is the inverse of
$\mathop{\tilde{g}}\nolimits
^{\nu\rho}_{(2)}$
 ($\mathop{\tilde{g}}\nolimits_{\mu\nu}\mathop{\tilde{g}}\nolimits
^{\nu\rho}_{(2)}=\delta^{\ \rho}_{\mu}$)
and $ \epsilon = \pm 1$ is the signature of the world-sheet metric.
For the $n=2$ gauge field, the constraint \decona\
 can be solved in terms of an unconstrained metric
tensor $g_{\mu \nu} $ as
$\tilde g^{\mu \nu}_{(2)}=\sqrt {\epsilon g}g^{\mu \nu} $, where
$g=  det
[g _{\mu \nu}]  $, so that the term
$\tilde g^{\mu \nu}_{(2)} \partial _ \mu \phi \partial _ \nu \phi$
becomes the standard minimal coupling to gravity.
If $\epsilon=1$, this metric has  Euclidean signature while if $\epsilon=-1$
the
signature is Lorentzian. Alternatively, the single constraint
$det (\tilde g_{(2)}^{\mu
\nu})=\epsilon$ on the three components of $\tilde g^{\mu \nu}_{(2)}$
can be solved in terms of two unconstrained functions $h_2(x), \bar h_2(x)$
which correspond to the two spin-two gauge fields of the previous section.
Similarly, the constraints on the spin-$n$ gauge field ${\tilde{g}}
^{\mu_{1}\mu_{2}...\mu_{n}}_{(n)} (x) $ can be
solved either in terms of tensor gauge fields satisfying algebraic trace
constraints, or
in terms of two unconstrained functions, which can be identified
with
  the gauge fields $h_n(x), \bar h_n(x)$ [\hegeoma].

  The full set of constraints are generated by the following
  constraint on $ \tilde F$:
 $$
det \left ({\partial ^2   \tilde F (x,y)\over \partial y_\mu
\partial y_ \nu} \right)=\epsilon
\eqn\decon$$
Expanding \decon\ in $y$ generates the full set of constraints.
This is the condition that $ \tilde F$ satisfies
 the real Monge-Amp\` ere equation [\mon].

The action \sis\ is invariant under the local $w_\infty$ transformations
 $$\delta\phi=\Lambda(x,\partial\phi)
\eqn\trfi$$
$$\eqalign{
\delta\mathop{\tilde{g}}\nolimits^{\mu_{1}\mu_{2}...\mu_{p}}_{(p)}&=\sum
_{m,n=2}^\infty
\delta_{m+n,p+2}\biggl[(m-1)\lambda^{(\mu_{1}\mu_{2}...}_{(m)}\partial
_{\nu}\mathop{\tilde{g}}\nolimits^{...\mu_{p})\nu}_{(n)}-(n-1)\mathop{\tilde
{g}}\nolimits^{\nu(\mu_{1}\mu_{2}...}_{(n)}\partial_{\nu}\lambda^{...\mu
_{p})}_{(m)} \cr &
+{(m-1)(n-1)\over p-1}\partial_{\nu}\left\lbrace\lambda^{\nu(\mu
_{1}\mu_{2}...}_{(m)}\mathop{\tilde{g}}\nolimits^{...\mu_{p})}_{(n)}-\mathop{\tilde
{g}}\nolimits^{\nu(\mu_{1}\mu_{2}...}_{(n)}\lambda^{...\mu_{p})}_{(m)}\right
\rbrace\biggr]
\cr}
\eqn\denvar
$$
where
$$\Lambda(x^{\mu},y_{\mu})=\sum^{\infty}_{n=2}\lambda^{\mu_{1}\mu
_{2}...\mu_{n-1}}_{(n)}\mathop{(x)y}\nolimits_{\mu_{1}}y_{\mu_{2}}...y_{\mu
_{n-1}}
\eqn\liss$$
for some infinitesimal symmetric tensor parameters $\lambda^{\mu_{1}\mu
_{2}...\mu_{n-1}}_{(n)}(x)$ which are required to satisfy the set of  algebraic
constraints generated  by expanding
  $$
\tilde F _{ \mu \nu} {\partial ^2   \over \partial y_\mu
\partial y_ \nu}  \Lambda (x,y) =0
\eqn\lamcon$$
 in $y$, where
$\tilde F _{ \mu \nu}(x,y)$ is the inverse of the matrix
${\partial ^2   \over \partial y_\mu
\partial y_ \nu}  \tilde F  (x,y)$
and can be written as
$$
\tilde F _{ \mu \nu}(x,y)=-   \epsilon _{ \mu \rho}
\epsilon _{\nu \sigma}
\tilde F ^{\rho \sigma}(x,y)
\eqn\qwerert$$
Using \decon, the constraint \lamcon\ can be rewritten (for
infinitesimal $ \Lambda$) as
 $$
det \left ({\partial ^2   \over \partial y_\mu
\partial y_ \nu} [\tilde F +\Lambda](x,y) \right)=\epsilon
\eqn\lamconoo$$
This constraint is  necessary for the transformations to
be a symmetry of the action [\geo,\hegeoma].
As will be seen in the next section,
the constraints \decon,\lamcon\ can be solved to give a theory which, in the
linearised limit, reproduces the linearised theory of the previous section.

The action is also invariant under the local symmetries
with parameters
$\alpha  ^{\mu_{1}\mu_{2}...\mu_{p}}_{(p,q)}
(x)$ for $q<p$ given by
$$\eqalign{
\delta {\tilde{g}} ^{\mu_{1}\mu_{2}...\mu_{p}}_{(p)}
=&
\alpha  ^{\mu_{1}\mu_{2}...\mu_{p}}_{(p,q)}
\cr
\delta {\tilde{g}} ^{\mu_{1}\mu_{2}...\mu_{q}}_{(q)}
=&-{q \over p}\alpha  ^{\mu_{1}\mu_{2}...\mu_{q}\mu_{q+1} \dots
\mu_{p}}_{(p,q)} y_{\mu_{q+1}} y_{\mu_{q+2}} \dots y_{\mu_{p}}
\cr}
\eqn\tejeltj$$
with all other fields
inert. These are the analogues of the \lq Stuckelberg' symmetries of
[\pop] and reflect the reducibility of the one-boson realisation of
$w_\infty$.
Nevertheless, most of the structure of the one-boson realisation
developed in this paper carries over immediately to multi-boson
realisations [\huprep] which are non-trivial and do not have Stuckelberg
symmetries.
 For further discussion, see
[\geo,\hegeoma,\hegeomb,\wevoo,\huprep].

\chapter{The Solution of the Constraints}

The constraint \decon\ can be given
 the following geometrical interpretation
[\geo,\hegeoma]. Let $\zeta_\mu, \bar \zeta _{\bar\mu}$ ($\mu =1,2$) be complex
coordinates on $\IR ^4$. Then, for each $x^\mu$, a solution $\tilde F(x,y)$ of
\decon\ can be used to define a function $K_x(\zeta , \bar \zeta)$ on $\IR ^4$
by $$K_x(\zeta , \bar \zeta)= \tilde F(x^\mu, \zeta_\mu +\bar \zeta _\mu)
\eqn\kis$$
For each $x$, $K_x$ can be viewed as the K\" ahler potential for a K\" ahler
metric $G^{\mu \bar \nu}= \partial ^2 K_x / \partial \zeta _\mu \partial \bar
\zeta _{\bar \nu}$ on $\IR ^4$. As a result of \decon, each $K_x$ satisfies the
Monge-Amp\` ere equation $det (G^{\mu \bar \nu})=\epsilon$
 and so the corresponding metric is K\" ahler and
Ricci-flat, which implies that the curvature tensor is either self-dual
or anti-self-dual. In the Euclidean case ($\epsilon=1$),
 the metric has signature $(4,0)$
and is hyperk\" ahler with $SU(2)$ holonomy, while in the Lorentzian case
($\epsilon=-1$)
the metric has signature
$(2,2)$ and holonomy $SU(1,1)$. As the K\" ahler potential is independent of
the imaginary part of $\zeta _\mu$, the metric has two
commuting (triholomorphic) Killing vectors, given by $i(\partial / \partial
\zeta _\mu - \partial / \partial \bar \zeta _{\bar \mu})$.
Thus the lagrangian $\tilde F(x,y)$ corresponds to a two-parameter family
of K\" ahler potentials $K_{x^\mu}$ for (anti-) self-dual geometries on $\IR
^4$
with two Killing vectors.
The parameter constraint \lamcon\
  implies that $\tilde F +\Lambda$ is  also   a K\" ahler
potential for a hyperk\" ahler metric with two Killing vectors.

Two solutions of the constraint  \decon\ were discussed in [\geo,\hegeoma]
 and both are
related to twistor transforms.
The first is the Legendre transform solution of
 [\hit]. Writing
 $y_{1}=\zeta$, $ y_{2}=\xi$,
$\tilde{F}(x^\mu,\zeta,\xi)$ can be written as the Legendre transform with
respect to $\zeta$ of some $\H$, so that
$$\tilde{F}(x,\zeta,\xi)=\pi\zeta -\H(x,\pi,\xi)
\eqn\leg$$
where the equation
$${\partial \H\over\partial\pi}=\zeta\ \eqn\treled$$
gives $\pi$ implicitly as a function of $x,\zeta,\xi$.
Taking the Legendre transform has the remarkable property of replacing the
complicated non-linear equation \decon\ with the
Laplace equation [\hit]
 $${\partial^{2}\H\over\partial\pi^{2}}+\epsilon {\partial^{2}\H\over\partial
\xi^{2}}=0
\eqn\hlap$$
and the general solution of this is
$$\H=f(x,\pi+\sqrt{- \epsilon} \xi)\ +\bar f(x,\pi-\sqrt {- \epsilon} \xi)
\eqn\hiss$$
where $f,\bar f$ are
  arbitrary independent real functions if $\epsilon =-1$ and are complex
conjugate functions if $\epsilon =1$. Then the general solution of \decon\
is  the Legendre transform \leg,\treled,\hiss\   and
  the action   can be given in the first order form
$$S=\ix\tilde{F}(x,y)=\ix \left[\pi \partial _\tau \phi -
 f(x^{\mu},\pi
+\partial _ \sigma \phi )+\bar f(x^{\mu},\pi-\partial _ \sigma \phi)
 \right]
\eqn\firs$$
where $\tau = x^1$ and $\sigma = -\sqrt{- \epsilon} x^2$.
The field equation for the auxiliary field $\pi$ is \treled\ and this can be
used in principle to eliminate $\pi$ from the action, but it will not be
possible to solve the equation \treled\   explicitly   in general. The
constraints \lamcon\
can be solved similarly.
 Expanding
the functions $f,\bar f$ gives the Hamiltonian form of the
$w_\infty$ action [\mik]
$$
S=\ix \left(
\pi \partial _\tau \phi- \sum _{n=2}^\infty {1 \over n}
\left[
h_n(\pi
+\partial _ \sigma \phi)^n +\bar h_n(\pi
-\partial _ \sigma \phi)^n
\right] \right)
\eqn\mikac$$

A related solution [\geo,\hegeoma] that involves
transforming with respect to both components of $y_\mu$ and maintains
Lorentz covariance was suggested by the results of [\van] and their
generalisation [\pop,\hue].
It will be useful to introduce a background
\lq metric'
 $\weta ^{\mu \nu}(x)$
on the cotangent space,
satisfying the constraint
$$
det(\weta ^{\mu \nu})= \epsilon
\eqn\wetcon$$
This constraint can be solved
in terms of an unconstrained background \lq metric'
$\wet  ^{\mu \nu}$
by
$$\weta ^{\mu \nu}= [\epsilon
det(\wet  ^{\mu \nu})]^{-1/2} \wet  ^{\mu \nu}
\eqn\wetsol$$
(Note that $\weta ^{\mu \nu}$ only determines
$\wet  ^{\mu \nu}$ up to a Weyl rescaling.)
 In [\geo,\hegeoma], this \lq metric'  was chosen to be the flat
metric
$$\weta ^{\mu \nu}(x)=\wet
 ^{\mu \nu}(x)= \eta ^{\mu \nu}\eqn\fla$$
but here it will be
useful to allow a more general choice. Different choices
will give equivalent results, but a judicious choice in which
$\weta ^{\mu \nu}$
transforms as a tensor density
 will be seen
later to lead to manifestly covariant results.

   $\tilde F(x^\mu ,y_\nu)$ is written
as a transform of a function $H(x^\mu , \pi^\nu)$ as follows:
 $$\tilde{F}(x
^{\mu},y_{\nu})=2\pi^{\mu}y_{\mu}-{1\over2}\weta^{\mu
\nu}y_{\mu}y_{\nu}-2H(x,\pi)
\eqn\fpi$$
where the equation
$$y_{\mu}={\partial H\over\partial\pi^{\mu}}
\eqn\yis$$
implicitly determines
$\pi^{\mu}=\pi^{\mu}(x^{\nu},y_{\rho})$.
The transform again linearises \decon\ and $\tilde F$   satisfies \decon\ if
and only if its transform $H$ satisfies
$${1 \over 2}
\weta^{\mu\nu}{\partial^{2}H\over\partial\pi^{\mu}\partial \pi^{\nu
}}= 1 \eqn\poi$$
and $\weta ^{\mu \nu}$ satisfies \wetcon.
This is true for any \lq background
 metric' $\weta ^{\mu \nu}$ satisfying \wetcon.

It will be useful to introduce a zweibein ${e_\mu} ^a$ $(a=1,2)$
such that $\wet ^{\mu \nu}
=\eta^{ab}{e_\mu} ^a{e_\nu} ^b=2 {e_{(\mu}} ^+{e_{\nu)}} ^-$
with
$e_\mu ^ \pm = { 1 \over \sqrt 2} (e_\mu ^1  \pm e_\mu ^2)$,
and define $\pi^1,\pi^2 $ by $\pi^a=e {e^a}_\mu \pi ^\mu$ where
$e=det({e^a}_\mu)$, together with the null coordinates
 $\pi ^\pm
= {1 \over  \sqrt 2}(\pi ^1 \pm \sqrt{-\epsilon} \pi ^2)$
which are independent
real coordinates for Lorentzian signature ($ \epsilon=-1$) and are complex
conjugate  coordinates for Euclidean signature $(\epsilon=1)$.
The general solution of \poi\ can now be written as
$$H=e\left[\pi^{+}\pi^{-}+f(x,\pi^{+})+\bar{f}(x,\pi^{-})
\right]
\eqn\hisp$$
(where $\bar f = f ^*$ if $ \epsilon =1$, but $f, \bar f$ are independent real
functions
 if $ \epsilon=-1$).
This solution
can be used to write the action  $$
S=
\int d^{2}x\left(2\pi^{\mu}y_{\mu}-\weta
_{\mu\nu}\pi^{\mu}\pi^{\nu}-{1\over2}\weta^{\mu\nu}y_{\mu}y_{\nu}-2ef(x,\pi
^{+})-2e\bar{f}(x,\pi^{-})\right)
\eqn\stac$$
where $\weta _{\mu \nu}$ is the inverse of $\weta ^{\mu \nu}$.
The field equation for $\pi^\mu $ is \yis, and using this to substitute for
$\pi$ gives the action \sis\ subject to the constraint \decon\ (details
are given in the appendix).
Alternatively, expanding the functions $f,\bar f$ as
$$
f = \sum _{n=2}^\infty { 1 \over n}h_n(x)(\pi ^+)^n
,\qquad
\bar f = \sum  _{n=2}^\infty { 1 \over n}\bar h_n(x)(\pi^-)^n
 \eqn\fofo$$
gives precisely the form of the action given in [\pop], following the
approach of [\van]. The parameter constraint \lamcon\ is solved similarly, and
the solutions can be used to write the symmetries of \stac\ in a form
similar to that
given in [\pop]. For example, the variation of $\phi$
given by $ \delta \phi = \Lambda (x,y)$ with $y_ \mu= \partial _ \mu\phi$
becomes
$$
\delta \phi= \Lambda(x^\mu , \pi^\nu)=\Lambda(x  , y(\pi))
\eqn\rtye$$
where $y( \pi)$ is found by solving ${ \partial \tilde F / \partial
y _ \mu}= 2 \pi^\mu - \weta ^{\mu \nu}y_ \nu$.
The constraint \lamcon\ on $\Lambda(x,y)$ then becomes the
following  simple
linear Laplace equation constraint on $\Lambda( x,\pi)$:
$$\weta ^{\mu \nu}{\partial ^2 \Lambda \over
 \partial \pi ^\mu \partial \pi ^\nu }=0
\eqn\lais$$
The calculation leading to this result is
given in the appendix.

The part of  $H$ quadratic in $\pi$
is
$$\half (\weta^{\mu \nu} \pi_\mu \pi _\nu + h_2 (\pi^+)^2
+ \bar h_2 (\pi^-)^2)\eqn\pquad$$
 and the terms involving $h_2, \bar h_2$
consist  of a background part ($\weta^{\mu \nu} $) and
a perturbation involving $h_2, \bar h _2$.
Different choices of  $\weta^{\mu \nu} $ correspond to
expanding the full metric about different background metrics.
The action \stac\ is invariant under spin-two transformations
for any choice of background $\weta^{\mu \nu}$;
different choices  lead to different transformation
rules. For example, with the choice $ \weta^{\mu \nu}=\eta^{\mu \nu}$, the
action \stac\ becomes precisely that of [\pop] and the transformations are
those given in [\pop]. If instead $\weta^{\mu \nu}$ is chosen to be a tensor
density transforming as
$$
\delta \weta^{\mu \nu}= k^ \rho \partial _ \rho \weta^{\mu \nu}
-2 \weta^{\rho (\mu }\partial _\rho k^{\nu)} +
  \weta^{\mu \nu} \partial _ \rho k^ \rho
\eqn\ghshg$$
under spin-two transformations with parameter $k^\mu= \lambda _{(2)}^\mu$
and $\pi^\mu$ is also chosen to be a tensor density, then
$\wet^{\mu \nu}$ transforms as a tensor, $\pi^a$ is a coordinate scalar
the first three
terms
in \stac\ are manifestly coordinate invariant and the remaining terms will be
invariant if the gauge fields $h_n, \bar h_n$ are chosen to
transform as scalars under coordinate transformations and as spin $n$
tensors under two-dimensional local Lorentz transformations.
Then the part of  $H$ quadratic in $\pi$
is given by \pquad\ and the terms involving $h_2, \bar h_2$
can be absorbed into a shift of $\weta^{\mu \nu} $. After this shift, the
action is given by \stac,\fofo,
in terms of  the
new shifted $\weta^{\mu \nu}$, which is again a tensor density, but now with
$$h_2=\bar h_2=0
\eqn\hisno$$
As
a result, $\tilde g^{\mu \nu}_{(2)}$, the spin-two gauge field in the expansion
\fis\ of $\tilde F$,  is now given by
$$
\tilde g^{\mu \nu}_{(2)}=\weta^{\mu \nu}
\eqn\choice$$
   giving a
formulation similar to that of [\van]. This has the advantage that the
invariance under diffeomorphisms is manifest, although the
shift of variables leads to a formulation in which the spin-two gauge fields
are no longer on an equal footing with the higher-spin ones.

For each $x^\mu$, the variables $h_n(x), \bar h_n (x)$ parameterise the
space of K\" ahler potentials $K_x$ (given by \kis) which are solutions of
the Monge-Amp\` ere equation, so that the $h_n, \bar h_n$ can be
taken to be the moduli of self-dual metrics on $ \IR ^4$ with two
commuting Killing vectors. For the family of geometries labelled by the
world-sheet coordinates $x^\mu$, the moduli become
functions $h_n(x), \bar h_n (x)$
of $x^\mu$ and these functions are interpreted as the gauge fields
of $w_ \infty$ gravity.

\chapter {Non-Linear $\W_N$ Gravity}

{}From the discussion in the introduction, the linearised action for
$\W_N$ gravity (\ie\ the action to linear order in the gauge fields) is
 an $N$'th order polynomial in $\partial _\mu \phi$ given by \noeth.
  However, the full
non-linear action is non-polynomial in the gauge fields and in
$\partial _\mu \phi$, but the
coefficient of $(\partial   \phi)^n$ for $n>N$ is a
polynomial function of the finite number of fundamental gauge fields  that
occur in the linearised action. The simplest way in which this might come
about would be if the action   were given by \sis,\fis\ and $\tilde F$
satisfies a constraint of the form
$${\partial^{N+1}\tilde{F}\over\partial y_{\mu
_{1}}\partial y_{\mu_{2}}...\partial y_{\mu_{N+1}}}=0+O(\tilde F^2)
\eqn\noo$$
where the right hand side is non-linear in  $\tilde F$ and its derivatives,
and depends only on derivatives of $ \tilde F$ of order $N$ or less.
It will be shown in this section that this is indeed the case; the action for
$\W_N$ gravity is given by \sis\ where
$\tilde F$ satisfies \decon\ and \noo, and the right hand
side of \noo\ will be
given explicitly. Just as the non-linear constraint \decon\ had an interesting
geometric interpretation, it might be expected that the non-linear form
of \noo\ should also be of geometric interest. It is essential that \noo\
should be consistent with the Monge-Amp\` ere   constraint \decon.

In the last section, the action for $w_\infty$ gravity was given in terms of
a function $\H(x^\mu,\pi,\xi)$ satisfying \hlap\ or a function $H(x^\mu
,\pi^\mu)$ satisfying \poi.
It follows from the results of [\van-\hue,\mik] that
these same actions can be used for $\W_N$
gravity provided that
the functions $\H$ or $H$ are restricted to be $N$'th order
polynomials in $\pi$ or $\pi^\mu$. The canonical first
order form of the $\W_N$ gravity action is then given    by \firs\ where
$\H$ \hiss\ satisfies \hlap\ and
$${\partial^{N+1}\H\over\partial\pi^{N+1}}=0
\eqn \nop$$
so that expanding the functions $f, \bar f$ gives the
action \mikac, but with the summation now running from $n=2$ to $n=N$ [\mik]
so that there are only a finite number of gauge fields
$h_n,\bar h_n$ where $n=2,3,\dots ,N$.

Similarly, the covariant first order form of the action is given by
\stac,\fofo\ where $H$ satisifies \poi\ and
$${\partial^{N+1}H\over\partial\pi^{\mu_{1}}\partial\pi^{\mu_{2}}...\partial
\pi^{\mu_{N+1}}}=0
\eqn\nog$$
so that $H$ \hisp\ is given by \fofo, with the summation   running
from $n=2$ to $n=N$ [\van-\hue].
Again, this leaves  a finite set of gauge fields,
$h_n,\bar h_n$ where $n=2,3,\dots ,N$
for $\W_N$ gravity,
 in agreement with the linearised
analysis.

It is remarkable that the constraints  defining $\W_N$ gravity -- \hlap,\nop\
or \poi,\nog\ -- are simple linear equations when written in terms of the $\pi$
variables. This can be understood in terms of the relation [\hit] between the
transform from $\tilde F(x,y)$ to $\H(x^\mu,\xi,\pi)$ or $H(x^\mu,\pi^\mu)$
and the Penrose transform, which translates the condition that a
geometry be self-dual into a linear twistor-space condition.
The Laplace equations \hlap,\poi\ become the Monge-Amp\` ere equation \decon\
when
written in terms of $\tilde F$ and it is this equation which characterises
$w_\infty$ gravity. The $\W_N$ condition, which is a complicated
non-linear constraint on $ \tilde  F$, becomes the simple linear constraint
\nop\ or \nog\ that the transform
$\H$ or $H$ is an $N$'th order polynomial in $\pi$.

\subsection{The Constraints on $ \tilde F$}

The equations \fpi,\yis\ give $\tilde F$ implicitly in terms of the function
$H$ and these can now be used to relate derivatives of
$\tilde F$ to those of $H$.
It will be useful to introduce the notation
$$H_{\mu_{1}\mu_{2}...\mu_{n}}={\partial^{n}H\over\partial\pi^{\mu
_{1}}\partial\pi^{\mu_{2}}...\partial\pi^{\mu_{n}}},\ \ \ \ \ \ F^{\mu
_{1}\mu_{2}....\mu_{n}}={\partial^{n}\tilde{F}\over\partial y_{\mu
_{1}}\partial y_{\mu_{2}}...\partial y_{\mu_{n}}}
\eqn\iuglg$$
and to define the inverse
 $H^{\mu\nu}$ of the \lq metric' $H_{\mu\nu}(x,\pi)$, so that
 $H^{\mu\nu} H_{\nu\rho}=\delta^{\mu}_{\
\rho}$.
Differentiating \fpi\ twice with respect to $y$ and using \yis\ and
 $${\partial\pi^{\mu}\over\partial y_{\nu}}=H^{\mu\nu}
\eqn\ettw$$
 gives
$$F^{\mu\nu}=-\weta^{\mu\nu}+2H^{\mu\nu}
\eqn\rtyhg$$
which can be used to give the \lq metric' $H_{\mu\nu}$ in terms of $\tilde F$
and the  metric $\weta^{\mu\nu}$:
 $$H_{\mu\nu}=2\left(\weta^{\mu\nu}+F^{\mu\nu}\right)^{-1}
\eqn\metis$$
Further differentiation yields
$$F^{\mu\nu\rho}=-2H^{\mu\alpha}H^{\nu\beta}H^{\rho\gamma}H_{\alpha
\beta\gamma}\eqn\rthf$$
$$F^{\mu\nu\rho\sigma}=-2H^{\mu\alpha}H^{\nu\beta}H^{\rho\gamma}
\mathop{H^{\sigma
\delta}H}\nolimits_{\alpha\beta\gamma\delta}+{3\over2}H_{\alpha\beta
}F^{\alpha(\mu\nu}F^{\rho\sigma)\beta}\eqn\fghg$$
$$\eqalign{
F^{\mu\nu\rho\sigma\tau}=&-2H^{\mu\alpha}H^{\nu\beta}H^{\rho\gamma
}\mathop{H^{\sigma\delta}H^{\tau\varepsilon}H}\nolimits_{\alpha\beta
\gamma\delta\varepsilon}+5H_{\alpha\beta}F^{\alpha(\mu\nu}F^{\rho
\sigma\tau)\beta}
\cr &
-{15\over4}H_{\alpha\beta}H_{\gamma\delta}F^{\alpha
(\mu\nu}F^{\rho\sigma|\gamma|}F^{\tau)\beta\delta}
\cr}
\eqn\thjury
$$
and it is straightforward to extend this to any number of derivatives
(see
the appendix for details).

Consider first $\W_3$ gravity. For $N=3$, the equation \nog\ becomes
$$H_{\alpha\beta\gamma\delta}=0\eqn\krhy$$
and using this \fghg\ becomes
$$F^{\mu\nu\rho\sigma}={3\over2}H_{\alpha\beta}F^{\alpha(\mu\nu}F^{\rho
\sigma)\beta}
\eqn\eryh$$
or, using \metis,
$$F^{\mu\nu\rho\sigma}=3\left(\weta^{\alpha\beta}+F^{\alpha\beta}\right)^{-1}F^{\alpha
(\mu\nu}F^{\rho\sigma)\beta}
\eqn\thcon$$
This is the required extra constraint
for $\W_3$ gravity.
Thus the action  for $\W_3$ gravity is given by \sis,\fis, where $\tilde F$ is
a function satisfying the two constraints \decon\ and \thcon.

Similarly, for $\W_4$ gravity, $H_{\alpha\beta\gamma\delta\epsilon}=0$ and
\thjury\ becomes
$$F^{\mu\nu\rho\sigma\tau}=5H_{\alpha\beta}F^{\alpha(\mu\nu}F^{\rho
\sigma\tau)\beta}-{15\over4}H_{\alpha\beta}H_{\gamma\delta}F^{\alpha
(\mu\nu}F^{\rho\sigma|\gamma|}F^{\tau)\beta\delta}
\eqn\fcon$$
so that the $\W_4$ action is \sis\ where $\tilde F$   satisfies
  \decon\ and \fcon, and $H^{\mu\nu}$ is given in terms of $\tilde F$ by
\metis.
Similar results hold for all $N$. In each case, taking the transform
of the linear constraint \nog\ yields
 an equation of
the form  \noo, where the right hand side is constructed from the $n$'th
order derivatives $F^{\mu_1 \dots \mu_n}$ for $2<n\le N$ and from $H_{\mu
\nu}$.

Expanding $\tilde F$ in $\dm \phi$ \fis\ gives the coefficient of the
$n$-th order $\partial
_{\mu _1} \phi \dots \partial
_{\mu _n} \phi $ interaction, which is proportional to $\tilde g^
{\mu _1  \dots  \mu _n}_{(n)}$. The constraint \noo\ implies that for $n>N$,
the coefficient $\tilde g _{(n)}$ of the $n$-th order interaction
can be written in terms of the coefficients $\tilde g _{(m)}$
of the $m$-th order interactions for $2 \le m \le N$.  For $\W_3$, the
$n$-point vertex can be written in terms of 3-point vertices for $n>3$, so
that
$$\mathop{\tilde{g}}\nolimits_{(4)}^{\mu\nu\rho\sigma}=2\left(\weta
^{\alpha\beta}+\mathop{\tilde{g}}\nolimits_{(2)}^{\alpha\beta}\right
)^{-1}\mathop{\tilde{g}}\nolimits_{(3)}^{\alpha(\mu\nu}\mathop{\tilde
{g}}\nolimits_{(3)}^{\rho\sigma)\beta}
\eqn\fion$$
$$\mathop{\tilde{g}}\nolimits_{(5)}^{\mu\nu\rho\sigma\tau}=5\left
(\weta^{\alpha\beta}+\mathop{\tilde{g}}\nolimits_{(2)}^{\alpha\beta
}\right)^{-1}\left(\weta^{\gamma\delta}+\mathop{\tilde{g}}\nolimits
_{(2)}^{\gamma\delta}\right)^{-1}\mathop{\tilde{g}}\nolimits_{(3)}^{\alpha
(\mu\nu}\mathop{\tilde{g}}\nolimits_{(3)}^{\rho\sigma|\gamma|}\mathop{\tilde
{g}}\nolimits_{(3)}^{\tau)\beta\delta}
\eqn\fitw$$
etc, while for $\W_4$, all vertices can be written in terms of 3- and 4-point
vertices, e.g.
$$\eqalign{
\mathop{\tilde{g}}\nolimits_{(5)}^{\mu\nu\rho\sigma\tau}
=&5\left
(\weta^{\alpha\beta}+\mathop{\tilde{g}}\nolimits_{(2)}^{\alpha\beta
}\right)^{-1}\mathop{\tilde{g}}\nolimits_{(3)}^{\alpha(\mu\nu}\mathop{\tilde
{g}}\nolimits_{(4)}^{\rho\sigma\tau)\beta}
\cr &
-5\left(\weta^{\alpha\beta
}+\mathop{\tilde{g}}\nolimits_{(2)}^{\alpha\beta}\right)^{-1}\left
(\weta^{\gamma\delta}+\mathop{\tilde{g}}\nolimits_{(2)}^{\gamma\delta
}\right)^{-1}\mathop{\tilde{g}}\nolimits_{(3)}^{\alpha(\mu\nu}\mathop{\tilde
{g}}\nolimits_{(3)}^{\rho\sigma|\gamma|}\mathop{\tilde{g}}\nolimits
_{(3)}^{\tau)\beta\delta}
\cr}
\eqn\fith$$
These \lq factorisations'
can be illustrated in Feynman-style diagrams. \fion\ is depicted in fig. 1,
where the
\lq propagators' represent
contraction of indices using the metric $H_{\mu \nu}$.
Similarly, \fitw\ and \fith\ are depicted in figs. 2 and 3 respectively, where
the \lq summation over channels' is not shown explicitly.

\subsection{The Constraints on $\Lambda$}

{}From the linearised analysis, it is expected that
the $ \W_N$ gravity action should be invariant under transformations
under which
$$
\delta \phi= \Lambda (x, \partial \phi)
\eqn\ghupoi$$
where $\Lambda(x,y)$ is of the form \liss\ and satisfies constraints whose
linearised forms are
$$
{ \partial ^2 \Lambda
\over \partial y_ \mu \partial y_ \nu } = 0 + \dots
\eqn\hgkh$$
and
$${\partial^{N } \Lambda \over\partial   y^{\mu_{1}}\partial
y^{\mu_{2}}...\partial
 y^{\mu_{N }}}=0 + \dots
\eqn\nogofgh$$
The full non-linear form of the constraint \hgkh\ is
given by \lamcon, while the non-linear form of
\nogofgh\ will give the parameters $
\lambda^{\mu_{1}\mu
_{2}...\mu_{n-1}}_{(n)}$
for $n>N$ in terms of the parameters
$
\lambda^{\mu_{1}\mu
_{2}...\mu_{m-1}}_{(m)}$ for $m \le N$ and the gauge fields,
so that the number of independent symmetries is the same as in
the linearised theory.

The full non-linear form of these constraints will now
be found by transforming the corresponding constraints in the
covariant first order form of the theory.
The covariant first order form of the $w_ \infty$ action, given by
\stac,\fofo\ where $H$ satisfies \poi, is invariant under transformations
given explicitly in [\pop] and which include
\rtye\
where  $\Lambda( x,\pi)$ satisfies
the constraint
\lais. It follows from the results of [\van-\hue] that
the truncation to $ \W_N$ gravity is obtained by imposing the constraint
\nog, so that $H$ is an $N$'th order polynomial in $\pi$,
together with the constraint
$${\partial^{N } \Lambda
\over\partial\pi^{\mu_{1}}\partial\pi^{\mu_{2}}...\partial
\pi^{\mu_{N }}}=0
\eqn\nogo$$
on $ \Lambda(x, \pi)$, so that $ \Lambda(x, \pi)$ is an $(N-1)$'th order
polynomial in $\pi$.

In addition, the constraints on the gauge fields given by
\nog\ or \thcon,\fcon\ etc are not invariant under the $ \Lambda$
transformations, but they become invariant if   the $ \Lambda$ transformations
are supplemented by
compensating \lq Stuckelberg' transformations, as in [\pop].

Now, using the chain rule and \ettw, it is straightforward to
express derivatives of $\Lambda (x, \pi)$ with respect to $\pi$
in terms of derivatives of $ \Lambda(x,y) $ with respect to $y$.
For example,
$$\eqalign{
{ \partial \Lambda \over \partial \pi ^ \mu}
&={ \partial \Lambda \over \partial y _ \alpha}H_{\alpha \mu}
\cr
{ \partial ^2 \Lambda \over \partial \pi ^ \mu \partial \pi ^ \nu}
&={ \partial ^2 \Lambda \over \partial y _ \alpha \partial y _ \beta
}H_{\alpha \mu}H_{\beta \nu}+
{ \partial \Lambda \over \partial y _ \alpha}H_{\alpha \mu \nu}
\cr
{ \partial ^3 \Lambda \over \partial \pi ^ \mu \partial \pi ^ \nu
\partial \pi ^ \rho}
&={ \partial ^3 \Lambda \over \partial y _ \alpha \partial y _ \beta
\partial y _ \gamma
}H_{\alpha \mu}H_{\beta \nu}  H_{\gamma \rho}
\cr &
\qquad +3{ \partial ^2 \Lambda \over \partial y _ \alpha \partial y _ \beta
}H_{\alpha (\mu \nu}H_{ \rho)\beta  }+
{ \partial \Lambda \over \partial y _ \alpha}H_{\alpha \mu \nu \rho}
\cr
{ \partial ^4 \Lambda \over \partial \pi ^ \mu \partial \pi ^ \nu
\partial \pi ^ \rho \partial \pi ^ \sigma}
&={ \partial ^4 \Lambda \over \partial y _ \alpha \partial y _ \beta
\partial y _ \gamma \partial y _ \delta
}H_{\alpha \mu}H_{\beta \nu}  H_{\gamma \rho}H_{\delta \sigma}
+
{ \partial \Lambda \over \partial y _ \alpha}H_{\alpha \mu \nu \rho \sigma}
\cr & \qquad+
6{ \partial ^3 \Lambda \over \partial y _ \alpha \partial y _ \beta
\partial y _ \gamma
}H_{\alpha (\mu \nu}H_{|\beta | \rho}  H_{ \sigma)\gamma }
\cr &
\qquad +
{ \partial ^2 \Lambda \over \partial y _ \alpha \partial y _ \beta
}\left[
3H_{\alpha (\mu \nu}H_{ \rho \sigma)\beta  }
+4 H_{\alpha (\mu \nu \rho}H_{ \sigma)\beta  }
\right]
\cr}
\eqn\lders$$

Consider first the case of $ \W_3$ gravity.
In this case, $H_{\mu \nu \rho \sigma}=0$ and the constraint on
$ \Lambda (x, \pi)$ given by
$${ \partial ^3 \Lambda \over \partial \pi ^ \mu \partial \pi ^ \nu
\partial \pi ^ \rho}
=0
\eqn\jghldfg$$
can be rewritten in terms of $\Lambda(x,y)$ and
$\tilde F(x,y)$ using \lders\ as
$$
{ \partial ^3 \Lambda \over \partial y _ \alpha \partial y _ \beta
\partial y _ \gamma
}= {3 \over 2}
{ \partial ^2 \Lambda \over \partial y _ \mu \partial y _\rho }{\delta_\rho}^
 {(
\gamma}  F^{\alpha \beta) \nu} H_{\mu \nu}
\eqn\lconthr$$
where $H_{\mu \nu}= 2( \weta^{\mu \nu}+F^{\mu \nu})^{-1}$ as before.
This constraint gives all of the parameters
$\lambda_{(n)}$ for $n=4,5,6, \dots$ in terms of the gauge fields, the
diffeomorphism
 parameter $\lambda_{(2)}^\mu$ and the spin-three parameter
 $\lambda_{(3)}^{\mu \nu}$ which satisfies the tracelessness constraint
 $(\tilde g_{(2)}^{ \mu \nu})^{-1}\lambda_{(3)}^{\mu \nu}=0$.
For example, for spin-four the constraint implies that
$$\lambda _{(4)}^{ \mu \nu \rho}= 2(\tilde h^{\alpha \beta}+
\tilde g^{\alpha
\beta}_{(2)} )^{-1}\lambda _{(3)}^{\alpha ( \mu  }
\tilde g^{ \nu \rho)
\beta}_{(3)}
\eqn\lamfcon$$

The $W_3$ constraint \thcon\ gives a sequence of constraints on the gauge
fields,
the first
two of which are given by \fion\ and \fitw.
These constraints are not invariant under the gauge field
transformations, which are now  given by \denvar, but with all the parameters
for  $\lambda _{(s)}^{ \mu \nu \dots}$
for $s>3$ given in terms of  the $s=2$ and $s=3$ parameters.
They do become invariant, however, if the spin two and spin three
transformations are supplemented by a compensating Stuckelberg transformation
of the form given by \tejeltj\ with $p=4,q=2$.
The transformations of $\tilde g_{(3)}^{ \mu \nu \rho}$ are unmodified,
but the spin-three transformation of $\tilde g_{(2)}^{ \mu \nu}$, which
previously was zero, now becomes modified to
$$\delta\mathop{\tilde{g}}\nolimits^{\mu\nu}_{(2)}={1\over2}A^{(\mu\nu\rho
\sigma)}y_{\rho}y_{\sigma}
\eqn\tutgug$$
where
$$\eqalign{
A^{\mu\nu\rho\sigma}=&{2\over3}\lambda^{\mu\nu}_{(3)}
\partial_{\alpha}\mathop{\tilde
{g}}\nolimits^{\rho\sigma\alpha}_{(3)}
-{10\over3}\mathop{\tilde{g}}\nolimits
^{\alpha\mu\nu}_{(3)}
\partial_{\alpha}\lambda _{(3)}^{\rho\sigma}+{4\over3}\mathop{\tilde
{g}}\nolimits^{\mu\nu\rho}_{(3)}
\partial_{\alpha}\lambda _{(3)}^{\sigma\alpha}+{4\over
3}\lambda _{(3)}^{\alpha\mu}\partial_{\alpha}\mathop{\tilde{g}}\nolimits
^{\nu\rho\sigma}_{(3)}
\cr &
+2G_{\alpha\beta}\mathop{\tilde{g}}\nolimits^{\gamma\mu}_{(2)}\mathop{\tilde
{g}}\nolimits^{\rho\sigma\beta}_{(3)}
\partial_{\gamma}\lambda _{(3)}^{\alpha\nu
}-\ 2G_{\alpha\beta}\mathop{\tilde{g}}\nolimits^{\gamma\mu}_{(2)}\lambda _{(3)}
^{\alpha\nu}\partial_{\gamma}\mathop{\tilde{g}}\nolimits^{\rho\sigma
\beta}_{(3)}
-{4\over3}_{\alpha\beta}\mathop{\tilde{g}}\nolimits^{\alpha
\mu\nu}_{(3)}
\mathop{\tilde{g}}\nolimits^{\sigma\beta}_{(2)}\partial_{\gamma}\lambda _{(3)}
^{\gamma\rho}
\cr &
-{2\over3}G_{\alpha\beta}\mathop{\tilde{g}}\nolimits^{\alpha\mu
\nu}_{(3)}
\mathop{\tilde{g}}\nolimits^{\rho\sigma}_{(2)}\partial_{\gamma}\lambda _{(3)}
^{\gamma\beta}+{1\over3}G_{\gamma\delta}\mathop{\tilde{g}}\nolimits
^{\rho\sigma}_{(2)}\partial_{\alpha}\left(\mathop{\tilde{g}}\nolimits^{\mu
\nu\delta}_{(3)}
\lambda _{(3)}^{\gamma\alpha}\right)
\cr &
+{2\over3}G_{\gamma\delta}\mathop{\tilde
{g}}\nolimits^{\rho\sigma}_{(2)}\partial_{\alpha}\left(\mathop
{\tilde{g}}\nolimits
^{\alpha\nu\delta}_{(3)}\lambda^{\gamma\mu}_{(3)}\right)
-{2\over3}G_{\alpha\beta}\mathop{\tilde{g}}\nolimits^{\alpha\mu
\nu}_{(3)}\left\lbrack\lambda^{\rho\sigma}_{(3
)}\partial_{\gamma}\mathop{\tilde
{g}}\nolimits^{\beta\gamma}_{(2)}
+2\lambda _{(3)}^{\rho\beta}\partial_{\gamma}\mathop{\tilde
{g}}\nolimits^{\sigma\gamma}_{(2)}\right\rbrack
\cr &
+2G_{\gamma\delta}\lambda _{(3)}^{\gamma\mu}\mathop{\tilde{g}}\nolimits
^{\nu\rho\delta}_{(3)}\partial_{\alpha}\mathop{\tilde{g}}\nolimits^{\sigma
\alpha}_{(2)}
+{2\over3}G_{\gamma\delta}\left\lbrack\lambda _{(3)}^
{\gamma\alpha}\mathop{\tilde
{g}}\nolimits^{\mu\nu\delta}_{(3)}+2\lambda _{(3)}^
{\gamma\mu}\mathop{\tilde{g}}
\nolimits
^{\alpha\nu\delta}_{(3)}
\right\rbrack\partial_{\alpha}\mathop{\tilde{g}}\nolimits
^{\rho\sigma}_{(2)}
\cr}
\eqn\mess$$
Here
$$G_{\mu\nu}=H_{\mu\nu}|_{y=0}=2\left(\mathop{\tilde{h}}\nolimits
^{\mu\nu}+\mathop{\tilde{g}}\nolimits^{\mu\nu}_{(2)}\right)^{-1}
\eqn\uopitr$$
This unpleasant form of the transformations simplifies dramatically
if we choose a general  $\mathop
{\tilde{h}}\nolimits
^{\mu\nu}$, and absorb $h_2 ,\bar h_2$ into field redefinitions, so that
$\mathop
{\tilde{h}}\nolimits
^{\mu\nu}=\mathop{\tilde{g}}\nolimits^{\mu\nu}_{(2)}$ and
 $G_{\mu \nu}=(\tilde
g^{\mu \nu})^{-1}$. Then the frame components $A^{abcd}$ of \mess\ are
given by
$$\eqalign{
A^{++++}=&2\lambda _{(3)}^{++}\nabla_{+}
\mathop{\tilde{g}}\nolimits^{+++}_{(3)}
-2\mathop{\tilde
{g}}\nolimits^{+++}_{(3)}\nabla_{+}\lambda _{(3)}^{++}
\cr
A^{----}=&2\lambda _{(3)}^{--}\nabla_{-}\mathop{\tilde{g}}\nolimits^{---}_{(3)}
-2\mathop{\tilde
{g}}\nolimits^{---}_{(3)}\nabla_{-}\lambda _{(3)}^{--}\cr
 A^{+++-}=&A^{++--}=A^{+---}=0
\cr}
\eqn\uoupto$$
In the chiral limit $\tilde g_{(3)}^{---}=0$, the transformation rules of
[\hu]
are   recovered.

Similarly, for $\W_4$ gravity, the constraint
${ \partial ^4 \Lambda \over \partial \pi ^ \mu \partial \pi ^ \nu
\partial \pi ^ \rho \partial \pi ^ \sigma}
=0$ leads to the constraint
$$
\eqalign{
\Lambda^{\alpha\beta\gamma\delta}
= & 3H_{\mu\nu}F^{\mu(\alpha\beta}\Lambda^{\gamma\delta)\nu}-{3\over
4}H_{\mu\rho}H_{\nu\sigma}F^{\mu(\alpha\beta}F^{\gamma\delta)\nu}\Lambda
^{\rho\sigma}
\cr &
+2H_{\mu\nu}F^{\mu(\alpha\beta\gamma}\Lambda^{\delta
)\nu}
-3H_{\mu\nu}H_{\rho\sigma}F^{\varpi\rho(\alpha}F^{\beta\gamma|\sigma
|}\Lambda^{\delta)\mu}\cr}
\eqn\uouo$$
where
$$\Lambda^{\mu_{1}\mu_{2}...\mu_{n}}\equiv{\partial^{n}\Lambda\over
\partial y_{\mu_{1}}\partial y_{\mu_{2}}...\partial y_{\mu_{n}}}\eqn\optup$$
The only independent parameters are
$\lambda_{(2)}^\mu$, $\lambda_{(3)}^{\mu \nu}$ and $\lambda_{(4)}^{\mu \nu
\rho}$, and these are subject to the constraints
$$\eqalign{
\tilde g_{(2)  \mu \nu} \lambda_{(3)}^{\mu \nu}&=0
\cr
\tilde g_{(2)  \mu \nu} \lambda_{(4)}^{\mu \nu \rho}&=
{2 \over 3} \tilde g_{(3)}^{ \mu \nu \rho}\lambda_{(3)}^{\sigma \tau}
\tilde g_{(2) \mu \sigma} \tilde g_{(2) \nu \tau}
\cr}\eqn\hkjsdgkjslkgj$$
where
$\tilde g_{(2)  \mu \nu}\equiv (\tilde g_{(2)}^{ \mu \nu})^{-1}$.
As in the $\W_3$ case, the transformations of the gauge fields are again
modified by compensating Stuckelberg-type transformations.

\subsection{The Geometry of the Constraints}

To attempt a geometric formulation of these results,
note that while the second derivative of $\tilde F$ defines a metric, the
fourth derivative is related to a   curvature, and the
$n$'th derivative is  related to the $(n-4)$'th covariant derivative of
the curvature. The $\W_3$ constraint \thcon\ can then be written
as  a constraint on the
curvature, while the  $\W_N$ constraint \noo\ becomes a constraint on the
$(N-3)$'th covariant derivative of
the curvature. One approach, motivated by that of section 3,
is to
introduce a second K\" ahler metric $\mathop{\hat{K}}\nolimits_{x}$ on $\IR ^4$
(for each $x^ \mu \in M$) given
in terms of the potential $K_{x}$ introduced in \kis\ by
$$\mathop{\hat{K}}\nolimits_{x}=K_{x}+\weta^{\alpha\bar{\beta}}\zeta
_{\alpha}\mathop{\bar{\zeta}}\nolimits_{\bar{\beta}}
\eqn\erter$$
The corresponding metric is given by
$$\mathop{\hat{G}}\nolimits^{\mu\bar{\nu}}=\weta^{\mu
\bar{\nu}} +G^{\mu\bar{\nu}}
\eqn\grfe$$
Then if $\tilde F$ satisfies the $\W_3$ constraint \thcon, the
curvature tensor for the metric \grfe\ satisfies
$$\mathop{\hat{R}}\nolimits^{\mu\bar{\nu}\rho\bar{\sigma}}={1\over
2}\mathop{\hat{G}}\nolimits_{\alpha\bar{\beta}}\left\lbrack
T^{\alpha
\mu\bar{\nu}}T^{\bar{\beta}\bar{\sigma}\rho}+T^{\alpha\mu\bar{\sigma
}}T^{\bar{\beta}\bar{\nu}\rho}+T^{\bar{\beta}\bar{\nu}\mu}T^{\alpha
\rho\bar{\sigma}}+T^{\bar{\beta}\bar{\sigma}\mu}T^{\alpha \rho \bar{\nu
}}\right\rbrack
\eqn\cur$$
where
$$T^{\mu\nu\bar{\rho}}={\partial^{3}\hat{K}\over\partial\zeta_{\mu
}\partial\zeta_{\nu}\partial\mathop{\bar{\zeta}}\nolimits_{\bar{\rho
}}},\ \ \ \ T^{\bar{\mu}\bar{\nu}\rho}={\partial^{3}\hat{K}\over\partial
\mathop{\bar{\zeta}}\nolimits_{\bar{\mu}}\partial\mathop{\bar{\zeta
}}\nolimits_{\bar{\nu}}\partial\zeta_{\rho}}
\eqn\tis$$
This is similar to, but distinct from, the constraint of special geometry
[\spec]. Note that \cur\ is not a covariant equation as the definitions \tis\
are only valid in the special coordinate system that occurs
naturally in
\W-gravity. However,
  tensor fields $T^{\mu\nu\bar{\rho}},T^{\bar{\mu}\bar{\nu}\rho}$
  can be defined by requiring them to be
  given by \tis\ in the special coordinate system and to transform
  covariantly, in which case the equation \cur\ becomes
  covariant, as in the case of
   special geometry  [\spec].
For $\W_N$, this generalises to give a constraint on the  $(N-3)$'th covariant
derivative of the curvature, which is given in terms of tensors that can each
be written in terms of  some higher order derivatives  of the K\" ahler
potential
in the special coordinate system.

For each $x^\mu$, the solutions to the constraints for $\W_N$ gravity are
parameterised by the $2(N-1)$ variables $h_n, \bar h_n$ for $2 \le n \le N$
which are then the coordinates for the
$2(N-1)$ dimensional
moduli space
for the  self-dual geometry satisfying the $\W_N$ constraint.
For the $x$-dependent family of solutions, the moduli become the
fields $h_n (x), \bar h_n(x)$ on the world-sheet.

  Further properties and generalisations of these actions will
  be given elsewhere.

\appendix

The background \lq metric'  $\weta _{\mu \nu}$,
 which satisfies
$
det(\weta ^{\mu \nu})= \epsilon
$,
  can be written
in terms of an unconstrained \lq metric'
$\wet  ^{\mu \nu}$
as
$$\weta ^{\mu \nu}= \sqrt h \wet  ^{\mu \nu}
\eqn\wetsola$$
where $h= \epsilon
det(\wet  ^{\mu \nu}) $.
Then $\tilde F$ can be written as
$$
\tilde F= \sqrt h \left(
2 \hat \pi ^ \mu y_\mu - {1 \over 2} h^{ \mu \nu}
y_\mu y_\nu - 2 \hh \right)
\eqn\erghyf$$
where
$$
\hat \pi ^ \mu = h^{-1/2} \pi ^ \mu, \qquad \hh = h^{-1/2}H
\eqn\tgjgh$$
It is useful to introduce a
 zweibein ${e_\mu} ^a$ $(a=1,2)$ (with inverse $e_\mu ^a$, and
$e=det({e_\mu}
^a)$) such that
 $$h ^{\mu \nu}
=  \eta^{ab}{e^\mu} _a{e^\nu} _b=2  {e^{(\mu}} _+{e^{\nu)}} _-\eqn\uouo$$
where $ \eta ^{ab}$ is the flat metric given by $diag(\epsilon,1)$,
and
$$e_\mu ^\pm = {1 \over \sqrt 2} \left( e_\mu ^1 \pm   \sqrt{- \epsilon}
e_\mu ^2 \right)
\eqn\rjhtiohw$$
We define flat null coordinates
$\pi^a= e^a_\mu \hat \pi ^\mu=e e^a_\mu \pi ^\mu$,
so that
$\pi ^\pm$
 are independent real coordinates if the
signature is Lorentzian ($ \epsilon=-1$),
and are complex conjugate coordinates $(\pi ^+ = (\pi^-)^*)$ if
the signature is Euclidean $(\epsilon=1)$.
In either case, the flat metric in the $\pi ^\pm$ coordinate
system is
$$\eta_{ab}=\left(\matrix{0&1\cr\noalign{\smallskip}1&0\cr}\right
)\eqn\etis$$

The Lagrangian $\tilde F(x,y)$ is given by a transform of a function
$H(x,\pi)$. For general $H(x,\pi)$, the second derivative
$$H_{\mu\nu}\equiv
{\partial ^2 H \over
\partial \pi ^\mu \partial \pi ^\nu }=
e^{-1} \hh _{\mu \nu}, \qquad
\hh _{\mu \nu} \equiv
{\partial ^2 \hh \over
\partial \hat \pi ^\mu \partial \hat \pi ^\nu }
\eqn\hmniso$$
can be written in terms of $\hh_{ab}$
$$H_{\mu\nu}=
e^{-1} \hh_{ab}{e_\mu} ^a{e_\nu} ^b, \qquad
\hh_{ab}\equiv
 {\partial ^2 \hh \over
\partial \pi ^a \partial \pi ^b }=
 \left(\matrix{h&a\cr\noalign{\smallskip}a&\bar{h}\cr
}\right)\eqn\hmnist$$
for some $a(x,\pi),h(x,\pi),\bar h(x,\pi)$.

The  constraint \poi\ becomes
$${1 \over 2} \eta^{ab}\hh_{ab}= \hh_{+-}= 1
\eqn\poiflat$$
 and the general solution of this is
$$H=e \hh= e\left[ \pi^{+}\pi^{-}+f(\pi^{+})+\bar{f}(\pi^{-}) \right]
\eqn\hissss$$
where dependence on $x^\mu$ has been suppressed.
Differentiating twice with respect to $\pi$ gives $ \hh_{ ab}$ which,
in the $\pi^\pm$ frame, takes the form
$$\hh_{ab}=\left(\matrix{h&1\cr\noalign{\smallskip}1&\bar{h}\cr
}\right)\eqn\hmnis$$
where
$$h=
{ \partial ^2 f\over \partial (\pi ^+)^2}
 , \qquad { \partial ^2 \bar f\over \partial (\pi ^-)^2}
\eqn\hfhf$$
and $a=1$.
It will be convenient to define
$$\Delta=1-h\bar{h}
\eqn\detis$$
In the $ \pi^\pm$ frame, the determinant of \hmnis\ is
$$det_{\pm}(\hh _{ab})=- \Delta=-1+h\bar{h}
\eqn\pirtrt
$$
 while in the $ \pi ^1, \pi^2$ frame
the determinant is
$$det (\hh_{ab})=- \epsilon
det_{\pm}(H_{ab})=\epsilon \Delta
\eqn\uiop
$$
(The sign changes result from the fact that the
Jacobian for the change of coordinates
from $ \pi ^1, \pi^2$
to $ \pi^\pm$ is $- \sqrt {-\epsilon }$.)
The inverse of \hmnis\ is
$$\hh^{ab}=e^{-1}{e^a}_\mu {e^b}_\nu
H^{\mu\nu}={1\over1-h\bar{h}}\left(\matrix{-\bar{h}&1\cr\noalign{\smallskip}1&-h\cr
}\right), \qquad H^{\mu \nu} \equiv
(H_{\mu \nu})^{-1}
\eqn\hinvis$$

{}From \rtyhg,\etis,\hmnis, the second derivative of $\tilde F$ is
$$F^{\mu\nu}=e{e^\mu} _a{e^\nu} _bF^{ab}\eqn\erter$$
where
$$
F^{ab}
=-\eta^{ab}+2 \hh
^{ab}={1\over1-h\bar{h}}\left(\matrix{-2\bar
{h}&1+h\bar{h}\cr\noalign{\smallskip}1+h\bar{h}&-2h\cr}\right)
\eqn\fisso$$
and the inverse of this matrix is
$$\left(F^{ab}\right)^{-1}={1\over1-h\bar{h}}\left(\matrix{2h&1+h\bar
{h}\cr\noalign{\smallskip}1+h\bar{h}&2\bar{h}\cr}\right)
\eqn\finv$$
The determinant of $F^{ab}$ is $-1$ in the $\pi ^ \pm$ coordinate
system and so is $\epsilon$ in the $\pi^1, \pi^2 $ coordinate system.
Using \erter, this implies
  that \decon\ is indeed satisfied.
Now, using \hinvis,\etis,
$$\hh^{ac}\eta_{cd}\hh^{db}={1\over(1-h\bar{h})^{2}}\left
(\matrix{-2\bar{h}&1+h\bar{h}\cr\noalign{\smallskip}1+h\bar{h}&-2h\cr
}\right)
\eqn\jojo$$
so that \fisso\ gives
$$F^{ab}=\Delta \hh^{ac}\eta_{cd}\hh^{db}
\eqn\fdel$$
and
$$ (F^{ab} )^{-1}=\Delta^{-1}\hh_{ac}\eta^{cd
 }\hh_{db}\eqn\fdes$$

The variation of $\phi$ is given by a function
$\Lambda (x, \pi ^\mu)$ by \rtye, where $\Lambda$ satisfies the constraint
\lais. This can be rewritten as
$$\eta ^{ab}{\partial ^2 \Lambda \over
 \partial \pi ^a \partial \pi ^b }=0
\eqn\laisab$$
Now the second relation in
\lders\ can be written as
 $${\partial ^2 \Lambda \over
 \partial \pi ^a \partial \pi ^b }={\partial ^2 \Lambda \over
 \partial y_c \partial y_d } \hh_{ac} \hh_{bd}
+{\partial   \Lambda \over
 \partial y_c   } {\partial ^3 \hh \over
\partial \pi ^a \partial \pi ^b \partial \pi ^c}
\eqn\tututu$$
$(y_a = e_a^\mu y_\mu)$ and taking the trace gives
$$\eta ^{ab}{\partial ^2 \Lambda \over
 \partial \pi ^a \partial \pi ^b }=\eta ^{ab}
{\partial ^2 \Lambda \over
 \partial y_c \partial y_d } \hh_{ac} \hh_{bd}
\eqn\yuyurigj$$
using \poiflat.
Then the constraint \laisab\ becomes
$$
(F^{ab})^{-1}{\partial ^2 \Lambda \over
 \partial y_a \partial y_b }=0
\eqn\yertrigj$$
using
\fdes, and this is equivalent to the parameter constraint
\lamcon.

We now turn to the identities satisfied by the
derivatives of $ \tilde F$.
{}From
$$\tilde{F}(x
^{\mu},y_{\nu})=2\pi^{\mu}y_{\mu}-{1\over2}\weta^{\mu
\nu}y_{\mu}y_{\nu}-2H(x,\pi)
\eqn\fpit$$
and
$$y_{\mu}={\partial H\over\partial\pi^{\mu}}
\eqn\yistt$$
it follows that
$$
F^\mu   \equiv {\partial \tilde  F \over \partial y_ \mu}
= 2 \pi^\mu -\weta^{\mu \nu}y_ \nu
\eqn\rturty$$
$$F^{\mu \nu}   \equiv{\partial ^2 \tilde  F \over \partial
y_ \mu \partial y_ \nu} = 2  {\partial  \pi^\mu \over \partial y_ \nu}
-\weta^{\mu \nu}  \eqn\utout$$
$$
F^{\mu \nu \rho}  \equiv{\partial ^3 \tilde  F \over \partial y_ \mu \partial
y_ \nu \partial y_ \rho} = 2  {\partial ^2 \pi^\mu \over \partial y_ \nu
\partial y_ \rho}
\eqn\yuyu$$
$$
F^{\mu _1 \dots \mu_n}   \equiv{\partial ^n \tilde  F \over \partial y_ {\mu
_1} \dots \partial y_ {\mu_n}} = 2  {\partial ^{n-1} \pi^{\mu_1} \over \partial
y_ {\mu_2} \dots \partial y_ {\mu_{n-1}}}
\eqn\nother$$
Then differentiating \yistt\ gives
$$
{\partial y_ \mu \over \partial \pi^ \nu}=
{\partial^2 H \over \partial \pi ^\mu \partial \pi ^ \nu} \equiv H_{\mu \nu}
\eqn\qweqw$$
and hence
$$
{\partial \pi^ \mu \over \partial y_ \nu}=(H_{\mu \nu})^{-1} \equiv H^{\mu \nu}
 \eqn\ryt$$
 Differentiating this gives
 $$\eqalign{
{\partial \pi^ \mu \over \partial y_ \nu \partial y_ \rho}&={\partial
\over \partial \pi^ \sigma} (H_{\mu \nu})^{-1} {\partial \pi^ \sigma \over
\partial y_   \rho}
\cr &
=- H^{\mu \alpha} H^{\nu \beta} H^{\rho \gamma}
{\partial^3 H \over \partial \pi ^\alpha \partial \pi ^ \beta \partial \pi ^
\gamma}
\cr}
 \eqn\ryton$$
Substituting this in \yuyu\ gives \rthf.
Differentiating \ryton\ gives
$$\eqalign{
{\partial \pi^ \mu \over \partial y_ \nu \partial y_ \rho
\partial y_  \sigma}&=-{\partial
\over \partial \pi^ \tau} \left(
H^{\mu \alpha} H^{\nu \beta} H^{\rho \gamma}
{\partial^3 H \over \partial \pi ^\alpha \partial \pi ^ \beta \partial \pi ^
\gamma}
\right)
 {\partial \pi^ \tau \over \partial y_   \sigma}
\cr &
=-H^{\mu \alpha} H^{\nu \beta} H^{\rho \gamma}H^{ \sigma \delta}
H_{\alpha \beta \gamma \delta}
+3H_{\kappa (\alpha \beta}H_{\gamma \delta) \lambda}H^{\kappa \lambda}
H^{\mu \alpha} H^{\nu \beta} H^{\rho \gamma} H^{\sigma \delta}
\cr}
 \eqn\rytono$$
and this leads to \fghg. It is straightforward to generalise
these relations to higher derivatives, and also to represent
the results graphically in figures similar to figs. 1,2,3 in which
$H_{\mu_1 \mu_2 \dots \mu_n}$ is represented as an
$n$-point vertex and $H^{ \mu \nu}$ is represented as a propagator.

\refout
\vfill
\eject

\FIG\fione{The four-point interaction in $\W_3$ gravity can
be written in terms of
three-point interactions. The diagram represents this factorisation,
with the symmetrization over the indices corresponding to the \lq sum over
channels'.}
\FIG\fitwo{This diagram represents the factorisation of the
five-point interaction into three-point interactions in $\W_3$ gravity, with
the summation over channels suppressed.}
\FIG\fithr{This diagram represents the factorisation of the
five-point interaction into three-point
and four-point interactions in $\W_4$ gravity, with
the summation over channels suppressed.}

\figout
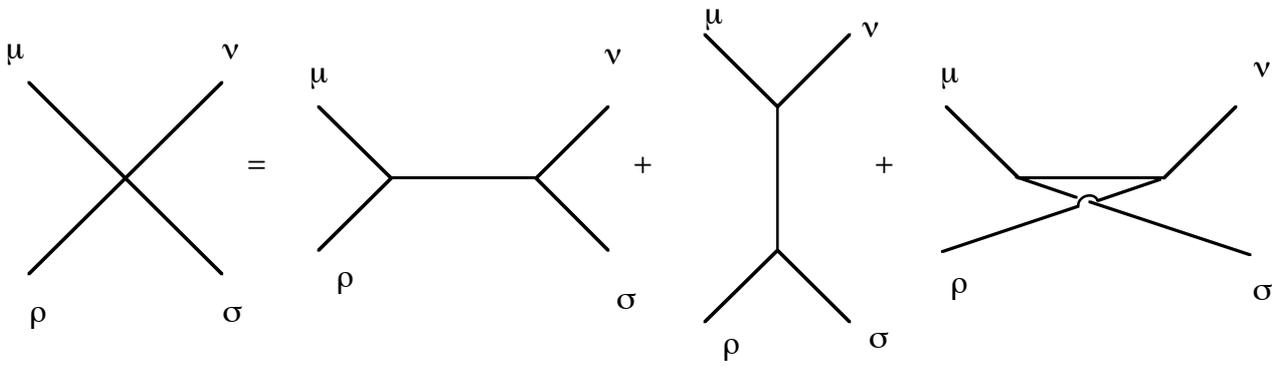
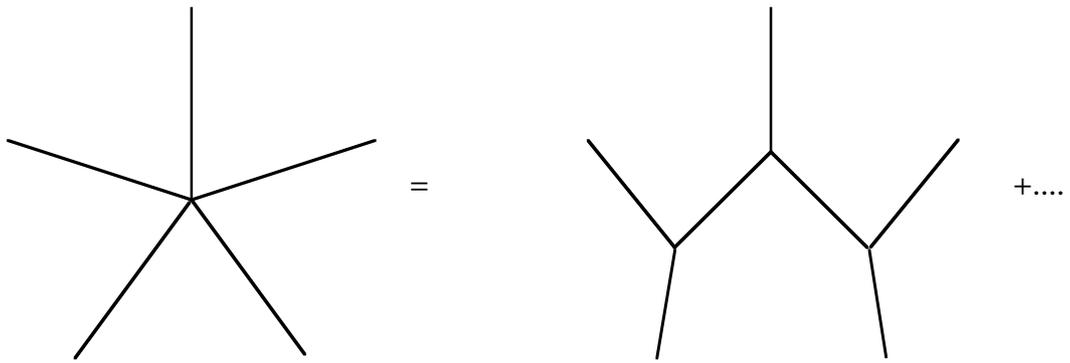
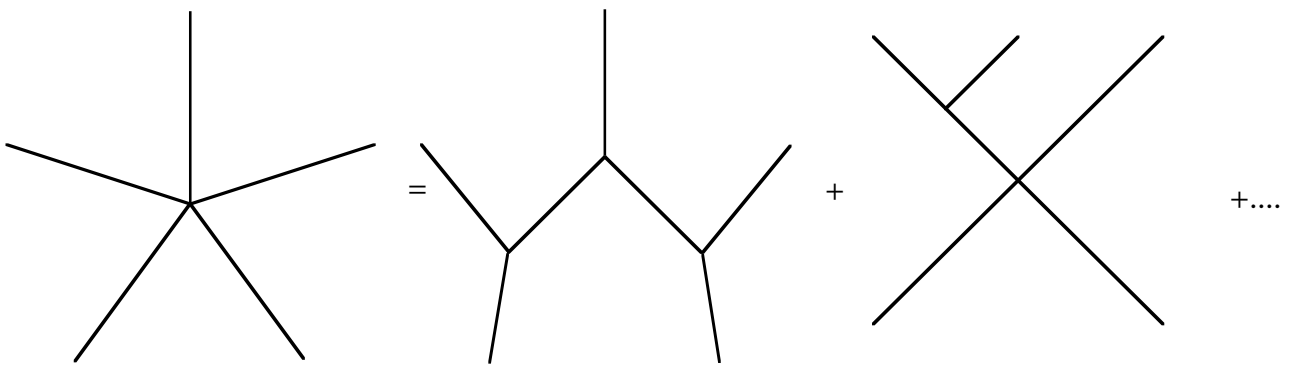
\end